\documentclass[twocolumn,showpacs,preprintnumbers,amsmath,amssymb,floatfix]{revtex4}

\usepackage{amsmath}
\usepackage{amsfonts}
\usepackage{graphicx}
\usepackage{dcolumn}
\usepackage{subfigure}

\newcommand{\ket}[1]{\ensuremath {|\: #1 \: \rangle}}
\newcommand{\bra}[1]{\ensuremath{\langle \: #1 \:|}}
\newcommand{\braket}[2]{\ensuremath{\langle \: #1 \: | \: #2 \: \rangle}}
\newcommand{\ketbra}[2]{\ensuremath{| \: #1 \:\rangle \langle \: #2 \:  |}}

\newcommand{\eref}[1]{(\ref{#1})}

\newcommand{\pd}[2]{\ensuremath{\frac{\partial #1}{\partial #2}  }}
\newcommand{\lapl}[2]{\ensuremath{\frac{\partial^2 #1}{\partial #2^2} }}
\newcommand{\llrr}[1]{\ensuremath{\left( #1\right)}}
\newcommand{\cd}{\cdot}

\begin{document}

\title{Quantum annealing and the Schr\"odinger-Langevin-Kostin equation}
\author{Diego de Falco}
\email{defalco@dsi.unimi.it}
\author{Dario Tamascelli}
\email{tamascelli@dsi.unimi.it}
\affiliation{%
Dipartimento di Scienze dell'Informazione, Universit\`a degli Studi di Milano, via Comelico 39/41, 20135 Milano, Italy}%

\date{June 27, 2008}

\begin{abstract}
We show, in the context of quantum combinatorial optimization, or quantum annealing, how  the nonlinear Schr\"odinger-Langevin-Kostin equation can dynamically drive the system toward its ground state. We illustrate, moreover, how a frictional force of Kostin type can prevent the appearance of genuinely quantum problems such as Bloch oscillations and Anderson localization which would hinder an exhaustive search.
\end{abstract}
\maketitle
\section{Introduction} \label{sec:intro}
The aim of combinatorial optimization is to find good approximations to the solution(s) of minimization problems. Many of the most famous algorithms currently used in this field \cite{papa98} were inspired by analogies with physical systems. Among them the most celebrated is \emph{Thermal simulated annealing} \cite{kirk83} proposed in 1983 by Kirkpatrick \emph{et al.}: the space of all admissible solutions is endowed with a potential profile dependent on the cost function associated to the optimization problem. The exploration of this space is represented by a temperature dependent random walk. An opportunely scheduled temperature lowering (annealing) stabilizes then the walk around a, hopefully global, minimum of the potential profile.\\
The \emph{Quantum annealing} approach to combinatorial optimization \cite{defa88,apo89}, instead, was originally suggested by the behaviour the stochastic process $q_\nu(t)$ associated \cite{albe77,eleu94} to the ground state $\phi_\nu$ of a Hamiltonian of the form:
\begin{equation} \label{eq:schr1}
H_\nu = -\frac{\nu^2}{2} \frac{\partial^2}{\partial x^2} + V(x),
\end{equation}
where the potential function $V$ encodes the cost  function to be minimized. The behavior of $q_\nu$ is characterized by long sojourns around the stable configurations, i.e. minima of $V(x)$, interrupted by rare large fluctuations which carry $q_\nu$ from one minimum to another:  $q_\nu$ in thus allowed to ``tunnel'' away from local minima to the global minimum of $V(x)$.  The deep analysis of the semiclassical limit performed in \cite{jona81} shows, indeed, that as $\nu \to 0^+$ \emph{``the process will behave much like a Markov chain whose state space is discrete and given by the stable configurations''}.\\
However, the ground state of $H_\nu$ is seldom exactly known and approximations are required. One of the earliest proposals in this direction, advanced in \cite{apo89} and applied in \cite{defa88}, was to construct an unnormalized approximation of $\phi_\nu(x)$ by acting on a suitably chosen initial condition $\phi_{trial}(x)$ with the Hamiltonian semigroup $\exp(-t H_\nu)$, namely by solving, with the initial condition $\phi_{trial}$, the \emph{imaginary time} Schr\"odinger equation. Similar ideas appear in the chemical physics literature \cite{anderson75} and,  with more specific reference to the optimization problems considered here, in \cite{piela89,amara93,finnila94}. Yet, the inability to autonomously construct the ground state process, without recourse to the unphysical step of imaginary time evolution, substantially detracts from what is otherwise a physical route to optimization by dynamical evolution toward the ground state.\\
In this note, encouraged by the progress in quantum annealing in the last twenty years, as reviewed for instance in \cite{santoro06,santoro08,das08}, by its close relationship with adiabatic quantum computation \cite{farhi01} and by proposals of its hardware implementation \cite{rose07}, we try to eliminate the above unphysical step: we try to implement, instead of imaginary time evolution, the idea of reaching the ground state with the help of viscous friction \cite{kostin72,kostin75,sanin07}.\\[3pt]
We first introduce the nonlinear Schr\"odinger-Langevin-Kostin (SLK) equation and illustrate, by means of examples on two toy models, how  a frictional force acts in the continuous case. Then we turn to quantum combinatorial optimization and show how dissipation can, in the discrete case, balance genuinely quantum effects, such as Bloch oscillations and Anderson localization, which can hinder the search of optimal solutions.
%
\section{Continuous case}
The  SLK equation is the analogue of the Heisenberg-Langevin equation and represents a quantum analogue of classical motion with
frictional force proportional to velocity \cite{kostin72,kostin75}; it can be seen as a rough analogue of the classical Drude-Lorentz model of Ohmic friction, i.e. an approximate description of the motion of a quantum particle through matter with inelastic scattering.\\
A solution $\psi(t,x) = \sqrt{\rho(t,x)} e^{i S(t,x)}$ of the equation:
\begin{align}\label{eq:slk}
 i \pd{\psi(t,x)}{t} & = -\frac{\nu^2}{2} \lapl{\psi(t,x)}{x} + V(x) \psi(t,x) + \nonumber \\
& +\beta \  S(t,x) \psi(t,x)
\end{align}
satisfies the inequality \mbox{$\frac{d}{dt} \bra{\psi(t)} H_\nu \ket{\psi(t)} \leq 0$} for \mbox{$\beta \geq 0$}, $H_\nu$ being the Hamiltonian \eref{eq:schr1}. What we will show below is how  the \emph{norm preserving} dissipative evolution described by \eref{eq:slk} can dynamically drive a suitable initial condition $\psi(0,x)$ toward the ground state $\phi_\nu(x)$ of  $H_\nu$.\\
{\bf Toy model 1:} Require
\begin{align*}
 \phi_\nu(x) & = c_+ \exp \llrr{-\frac{\llrr{x-a}^2}{4 \sigma_+^2}}+ c_-\exp \llrr{-\frac{\llrr{x+a}^2}{4 \sigma_-^2}} +\\
& + c_0 \exp  \llrr{-\frac{x^2}{4 \sigma_0^2}},
\end{align*}
(the parameters $c_{\pm,0}$ being chosen so that $\phi_\nu(x) >0$) to be the ground state of $H_\nu$ and to belong to the eigenvalue $E_\nu = 0$. The above two requirements determine the potential  $V(x) = \frac{\nu^2} {2 \phi_\nu(x)} \frac{d^2}{dx^2} \phi_\nu(x)$.
\begin{figure}[t]
 \centering
	\subfigure[]{\label{fig:tela2008g2a} \includegraphics[width=7cm]{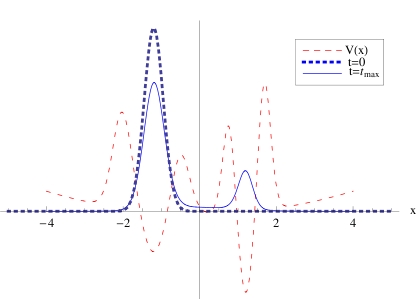}}
	\hspace{1cm}
	\subfigure[]{\label{fig:tela2008g1c}\includegraphics[width=7cm]{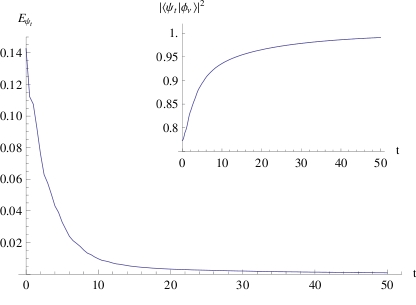}}
\caption{\label{fig:tela2008g2} (Color online) Frame (a): The probability density $\left | \psi(t,x) \right |^2$ at the initial time $t=0$ and final $t=t_{max}=50$. The potential $V(x)$ is drawn in arbitrary scale for expository purposes. Frame (b): The expected value $\bra{\psi(t)} H_\nu \ket{\psi(t)}$ of the Hamiltonian operator ($\nu=1,\ \beta=0.5$); in the inset: the overlap $\left | \braket{\psi(t)}{\phi_\nu} \right |^2$ between the ground state of $H_\nu$ and the state of the system at time $t$, for $0\leq t\leq t_{max} =50$.}
\end{figure}
\begin{figure}
 \centering
 \subfigure[]{\label{fig:tela2008g3a} \includegraphics[width=7cm]{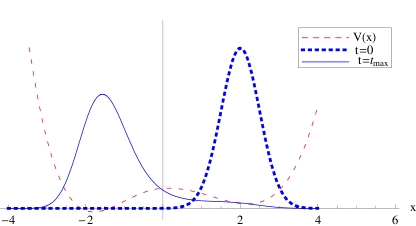}}\\
 \subfigure[]{\label{fig:tela2008g3cd} \includegraphics[width=7cm]{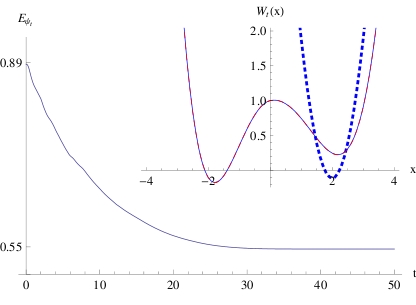}}
\caption{ \label{fig:tela2008g3} (Color online) Frame (a): The probability density $\left | \psi(t,x) \right |^2$ at the initial time $t=0$ and final $t=t_{max}=50$. The potential $V(x)$, corresponding to the choice $a_+ = 2.25,\ a_-=1.75,\ V_0=1,\ \delta = 0.1$ of the parameters of \eref{eq:potSantoro}, is drawn in arbitrary scale for expository purposes. Frame (b): The expected value $\bra{\psi(t)} H_\nu \ket{\psi(t)}$ of the Hamiltonian operator ($\nu=1,\ \beta=0.3$); in the inset: the potential profile $W(t,x)$ at the initial and final times: the potential profile at $t=t_{max} =50$ completely overalps $V(x)$.}
\end{figure}
Figure \ref{fig:tela2008g2} follows the evolution $\psi(t,x)$  of the initial condition $\psi(0,x) = \frac{\exp \llrr{- \frac{\llrr{x+a}^2}{4 \sigma_-^2} } }  {\llrr{2 \pi \sigma_-^2}^{1/4}}$ under \eref{eq:slk}  for a time interval $\llrr{0,t_{max}}$.
It shows that, as $\bra{\psi(t)}H_\nu \ket{\psi(t)}$ decreases with time, the ``vacuum overlap'' $\left |\braket{\psi(t)}{\phi_\nu} \right|^2$ approaches the value 1. While the state $\psi(t)$ approaches the ground state, some of the probability mass ``tunnels'' from the leftmost (local) to the rightmost (global) minimum.\\
We point out that this class of examples, where both the ground state wave function $\phi_\nu(x)$ and the ground state energy $E_\nu$ are known, allows also for the calibration of the numerical method. In our case we have used the built-in \texttt{NDSolve} resource of \emph{Wolfram Mathematica 6}.\\[3pt]
{\bf Toy model 2}: For the sake of comparison with classical literature on quantum annealing \cite{santoro06,santoro05}, we consider here a double-well potential of the form
\begin{equation} \label{eq:potSantoro}
 V(x) = \begin{cases}
        V_0 \frac{\llrr{x^2 - a_+^2}^2}{a_+^4} + \delta x, & \mbox{for $x \geq 0$} \\
	V_0 \frac{\llrr{x^2 - a_-^2}^2}{a_-^4} + \delta x, & \mbox{for $x < 0$}.
        \end{cases}
\end{equation}
As shown in figure \ref{fig:tela2008g3a}, for the parameters used there, the local minimum of the potential $V(x)$ is wider than the global one. We refer the reader to section 2.2 of \cite{santoro06} and to \cite{santoro05} for a discussion of the meaning of the parameters and for the presentation of numerical experiments comparable with ours. Here, we use this well known toy model as an example in which the ground state is \emph{unknown} and the dissipative dynamics of SLK type provides a method to find it.\\
While the initial condition evolves (figure \ref{fig:tela2008g3a}), the mean value of $H_\nu$ decreases as in figure \ref{fig:tela2008g3cd}. That $\psi(t_{max},x)$ is a good approximation of the ground state is shown by comparing, in the inset of figure \ref{fig:tela2008g3cd}, $V(x)$ with the potential $W(t,x)  = \frac{1} {2 \psi(t,x)} \frac{\partial^2}{\partial x^2}  \psi(t,x) + \bra{\psi(t)} H_\nu \ket{\psi(t)}$
evaluated at $t=t_{max}$, of which $\psi(t_{max},x)$ is the ground state belonging to the eigenvalue 0. Comparison of the two curves in the inset of figure \ref{fig:tela2008g3cd} is, furthermore, suggestive of a real-time version of Piela's \emph{method of deformation of the potential energy hypersurface} \cite{piela89}.\\
In the continuous case, then,  the SLK Hamiltonian enriches quantum annealing of what can be seen as a \emph{probability percolation}: while the state of the system converges toward the ground state, some probability mass tunnels from one minimum to another and, instead of tunneling back, as would happen in a reversible dynamics, remains there. This autonomous stabilization would represent an alternative to \emph{adiabatic quantum computation} \cite{farhi01}.
\section{Discrete case}
Most of the effort in Ref. \cite{apo89} went into making the intuition developed so far available in a context of \emph{combinatorial} optimization. In such a context the domain of the function $V$ to be minimized is a finite set $Q$ and the search of the minimum of $V$ is modeled on a graph $\llrr{Q,E}$, where the edges $e \in E$ describe the moves allowed in the search. For instance, in \cite{apo89}, $Q$ was taken to be the Boolean hypercube $Q_n = \{ -1,1 \}^n $, for some positive integer $n$, and an edge was placed between any two points in $Q_n$ separated by a unit Hamming distance.\\
In this note, we consider the much simpler instance in which $Q$ is, for some positive integer $s$, the finite set $\Lambda_s  = \{1,2,\ldots,s\}$ equipped with the set of edges \mbox{$E = \left \{ \{i,j\} : (i,j) \in \Lambda_s \times \Lambda_s \wedge |i-j|=1  \right \}.$} According to the general approach outlined in \cite{apo89}, this amounts to a search of the minimum of the function $V$ defined on $\Lambda_s$ by means of an interacting continuous time quantum walk \cite{defa06a} on $\Lambda_s$ governed by a Hamiltonian of the form  
\begin{align}
 h  &= - \frac{1}{2} \sum_{j=1}^{s-1} \ketbra{j+1}{j} +  \ketbra{j}{j+1}  + \nonumber \\
& +\sum_{j=1}^{s} V(j) \ketbra{j}{j}.
\end{align}
We wish to show here that the \emph{quantum} search outlined above can suffer from two, typically \emph{quantum}, problems, namely Bloch oscillations \cite{peschel98} and Anderson localization \cite{anderson58} and that a certain amount of ``viscous'' friction can provide some relief to both these problems.\\
On a finite box $\Lambda_s$, we consider the evolution of an initial condition of the form $c_k(x) = \sum_{x=1}^\epsilon \sqrt{\frac{2}{\epsilon+1}} \sin \llrr{\frac{k \pi x}{\epsilon+1}}$.
%
\begin{figure}
 \centering
{ \subfigure[]{\label{fig:tela2008g4a}\includegraphics[width=4cm]{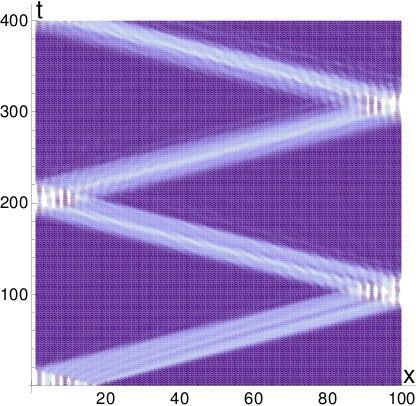}}}
\hspace{0.2cm}
{ \subfigure[]{\label{fig:tela2008g4b}\includegraphics[width=4cm]{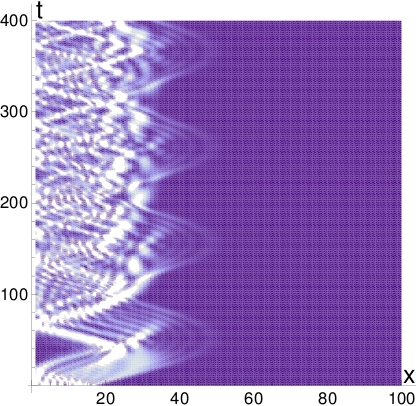}}}\\
{ \subfigure[]{\label{fig:tela2008g4c}\includegraphics[width=4cm]{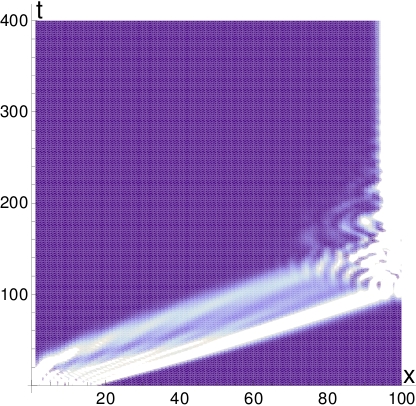}}}
\hspace{0.2cm}
{ \subfigure[]{\label{fig:tela2008g4d}\includegraphics[width=4cm]{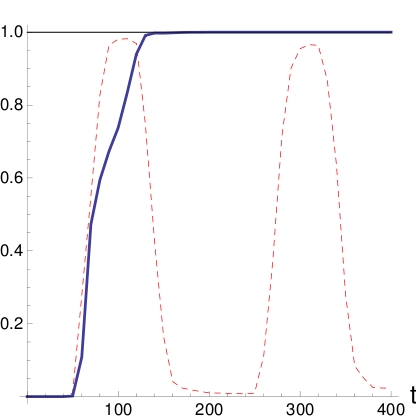}}}\\
\caption{ \label{fig:tela2008g4} (Color online) $s=100,\ \epsilon=17,\ k=(\epsilon-1)/2,\ g_0=2/s,\ 0\leq t \leq 4s$. Frame (a): free evolution ($g=0,\ \beta=0$); (b): $g=3 g_0,\ \beta=0$: Bloch oscillations prevent the wave packet from exploring most of the solution space; (c): $g=3 g_0,\ \beta=4 g_0$: Bloch oscillations no longer appear, thus allowing a complete exploration of the solution space and convergence to the ground state; (d) the probabilies, as a function of time, of reaching the $\delta = 2 \epsilon$ rightmost sites of $\Lambda_s$ in the free case (thin dashed line) and in the forced and damped case $g=3g_0,\ \beta=4 g_0$ (solid thick line).
\vspace{-0.5cm}
}
\end{figure}
\vspace{-0.1cm}
We refer the reader to section 5 of \cite{defa06a} for a motivation of this choice: suffice here to say that it describes a spatially well located wave packet that in \emph{the absence of any potential} moves back and forth, with speed close to 1, inside the box $\Lambda_s$, as in figure \ref{fig:tela2008g4a}. The effect on this \emph{ballistic} evolution of a linear potential $V(x) = -g x$ is shown in figure \ref{fig:tela2008g4b}. For $g=O(1/s)$, the peculiar energy-momentum relation $E(p) = 1-\cos p$, holding on a discrete lattice, determines Bloch oscillations that prevent the wave packet from approaching the point $x=s$ at which the minimum of the cost function $V(x)$ is located. Figure \ref{fig:tela2008g4b} is therefore a reminder of the fact that a \emph{greedy} quantum optimization driven by the cost function itself acting as a potential can be hindered by the fact that, on a lattice, increasing momentum $p$ can mean decreasing velocity $v(p)=\sin p$.\\[3pt]
We propose here to introduce a certain amount of viscous friction in the discrete Schr\"odinger equation, as a ``Kostin potential'' $K(t,x) \approx \beta \ S(t,x) = \beta \  Arg\left (  \psi(t,x) \right)$:
\begin{align}\label{eq:hamcompl}
 i \frac{ \partial \psi(t,x)}{\partial t} & = (h \ \psi)(t,x) + K(t,x) \psi(t,x) = \\
 & = -\frac{1}{2} \left (  \psi(t,x+1) + \psi(t,x-1)\right ) +  \nonumber \\
& + V(x) \psi(t,x) + K(t,x) \psi(t,x). \nonumber
\end{align}
The idea is that friction can prevent the momentum $p$ from crossing the first Brillouin zone and thus can prevent the velocity $\sin p$ from being inverted before the wave packet reaches the boundary of $\Lambda_s$. This unwanted inversion is illustrated in figure \ref{fig:tela2008g4b}, the effect on it of a suitable Kostin potential is shown in figure \ref{fig:tela2008g4c}.\\
As it is easy to check that, for $\psi(t)$ evolving \mbox{according to \eref{eq:hamcompl}, it is}
\begin{align}
 &\frac{d}{dt} \bra{\psi(t)} h \ket{\psi(t)}  = - \sum_{x=1}^{s-1} \sqrt{\rho(x+1) \rho(x)} \cd \\
 &\cd  \left(K(t,x+1) - K(t,x) \right)  \sin \left(S(t,x+1)- S(t,x) \right), \nonumber
\end{align}
the actual form of $K(t,x)$ that we adopt in order to achieve decrease of $\bra{\psi(t)} h \ket{\psi(t)}$ is \mbox{$K(t,x) = \beta \sum_{y=2}^x \sin \left( S(t,y) - S(t,y-1)\right)$}, with $\beta >0$.
%
%
%
%
\begin{figure}
 \centering
{ \subfigure[]{\label{fig:tela2008g5a}\includegraphics[width=4cm]{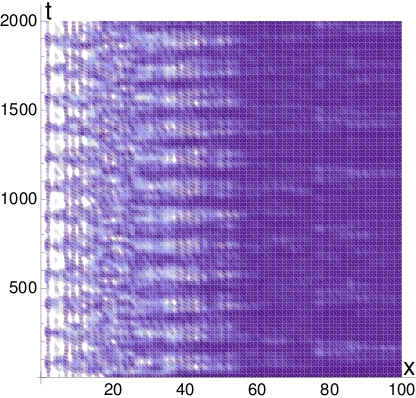}}}
\hspace{0.2cm}
{ \subfigure[]{\label{fig:tela2008g5b}\includegraphics[width=4cm]{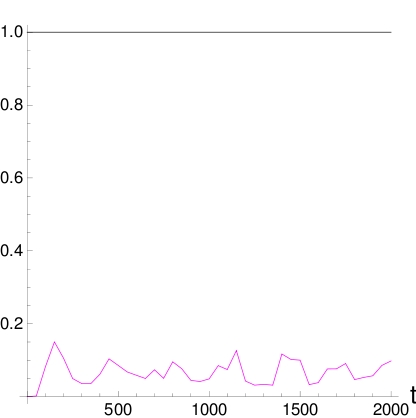}}}\\
{ \subfigure[]{\label{fig:tela2008g5c}\includegraphics[width=4cm]{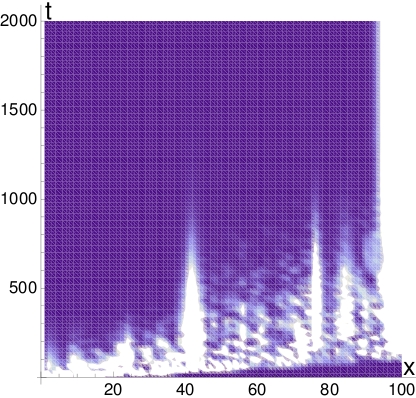}}}
\hspace{0.2cm}
{ \subfigure[]{\label{fig:tela2008g5d}\includegraphics[width=4cm]{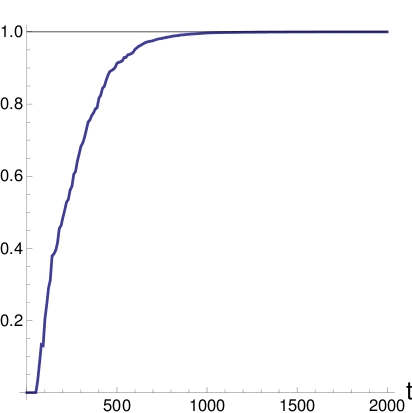}}}\\
\caption{ \label{fig:tela2008g5} (Color online) $s=100,\ \epsilon=17,\ k=(\epsilon-1)/2,\ g_0=2/s,\ \sigma_0 = \llrr{10/s}^{3/2},\ 0\leq t \leq 20s$. Frame (a): $g = 0,\ \beta=0,\ \sigma = 2 \sigma_0$: the wave packet gets confined in the first half of $\Lambda_s$; (b): for $g = 0,\ \beta=0,\ \sigma = 2 \sigma_0$ the probability of ever reaching the $\delta = 2 \epsilon$ rightmost sites is negligible (Anderson localization); (c): $g=3 g_0,\ \beta = 4 g_0, \sigma= 2 \sigma_0$: viscous friction allows the particle to drift to the right, by successive sojourns (the vertical strips) around successive minima of the Anderson potential; the ensuing slow transfer of the probability mass to the right of $\Lambda_s$ is shown in frame (d).
\vspace{-0.5cm}
}
\end{figure}
\\[3pt]
Figure \ref{fig:tela2008g5a} shows, instead, the effect, in the form of Anderson localization, of a random Gaussian potential of mean 0 and variance $\sigma^2$, acting independently on each site of $\Lambda_s$. The order of magnitude $\sigma_0 = \llrr{10/s}^{3/2}$ of the noise parameter $\sigma$ is suggested by a scaling \mbox{argument  \cite{malyshev04}.}\\[3pt]
Whereas the fact that friction can wipe out Bloch oscillations is well known \cite{sanin07}, the less well known fact that we show here is that the pseudo-ballistic motion shown in figure \ref{fig:tela2008g4c} is \emph{much more stable} than the truly inertial motion represented in figure \ref{fig:tela2008g4a} with respect to the onset of Anderson localization.\\[3pt]
%
\section{Conclusions and outlook}
As a final remark, we observe that the same framework developed in this note for the combinatorial optimization metaphor can be used, with minor changes, to describe an excitation travelling along a spin chain or a light pulse propagating through a waveguide lattice \cite{perets07}. We conjecture, therefore, that SLK dynamics can be exploited also in those fields. For example, we can, maybe, increase the fidelity of state transmission, in presence of imperfections, along a spin chain, by applying a ``tension'' at both ends of it \cite{feyn86} (see figure \ref{fig:tela2008g5d}). The sole convergence toward the ground state could, instead, find  applications in all-optical switching of light in waveguide arrays \cite{christo03}: the injected light pulse can be steered toward a given position by a suitable tuning of the thermal gradient which determines the potential profile of the lattice. Future work should be devoted to further investigation of this open research problems.
%

%
\end{document}